\documentclass[superscriptaddress, preprint, prb]{revtex4}

\usepackage[T1]{fontenc}                  
\usepackage[english]{babel}
\usepackage[latin1]{inputenc}
\usepackage{textcomp}                     
\usepackage{graphicx}
\usepackage[thinspace,thinqspace, Gray]{SIunits}
\usepackage{flafter}
\usepackage{amsmath}
\usepackage{subscript}

\begin{document}
\title{Tunneling magneto thermocurrent in CoFeB/MgO/CoFeB based magnetic tunnel junctions}
\date{\today}
\author{N. Liebing}
\affiliation{Physikalisch-Technische Bundesanstalt, Bundesallee 100, D-38116 Braunschweig, Germany}
\author{S. Serrano-Guisan}
\affiliation{Physikalisch-Technische Bundesanstalt, Bundesallee 100, D-38116 Braunschweig, Germany}
\affiliation{International Iberian Nanotechnology Laboratory, Av.~Mestre Jos\'{e} Veiga, 4715-330 Braga, Portugal}
\author{P. Krzysteczko}
\affiliation{Physikalisch-Technische Bundesanstalt, Bundesallee 100, D-38116 Braunschweig, Germany}
\author{K. Rott}
\author{G. Reiss}
\affiliation{University of Bielefeld Department of Physics Universit\"{a}tsstr.~25, D-33615 Bielefeld, Germany}
\author{J. Langer}
\author{B. Ocker}
\affiliation{Singulus AG, Hanauer Landstrasse 103, D-63796 Kahl am Main, Germany}
\author{H. W. Schumacher}
\affiliation{Physikalisch-Technische Bundesanstalt, Bundesallee 100, D-38116 Braunschweig, Germany}
\begin{abstract}
We study the tunneling magneto thermopower and tunneling magneto thermocurrent of CoFeB/MgO/CoFeB magnetic tunnel junctions (MTJ). The devices show a clear change of the thermoelectric properties upon reversal of the magnetisation of the CoFeB layers from parallel to the antiparallel orientation. When switching from parallel to antiparallel the thermopower increases by up to \unit{55}{\%} where as the thermocurrent drops by \unit{45}{\%}. These observations can be well explained by the Onsager relations taking into account the tunneling magneto resistance of the MTJ. These findings contrast previous studies on Al$_2$O$_3$ based MTJ systems, revealing tunneling magneto thermo power but no tunneling magneto thermocurrent.
\end{abstract}
\maketitle
Spin-caloritronics \cite{Bauer2010} is an emerging field combining thermoelectrics and spintronics properties in magnetic nanostructures. It enables, for instance, the generation of pure spin currents by thermal gradients~\cite{Uchida2011a,Kikkawa2013,Qu2013} or the thermally induced excitation of magnetization precession~\cite{Jia2011}. These phenomena are driven by the interplay of heat, charge and spin currents. They can path the way towards novel functionalities for future energy-efficient information and communication technology and thus have triggered worldwide research activities.

CoFeB/MgO/CoFeB magnetic tunnel junctions (MTJs) are presently among the most important spintronics material systems. Recently, a large tunneling magneto thermopower (TMTP) - or tunneling magneto Seebeck effect - up to \unit{90}{\%} has been observed in such MTJs~\cite{Liebing2011, Liebing2012,Lin2012,Walter2011b}. There, a thermal gradient across the tunnel barrier induces a thermovoltage \textit{V}\textsubscript{TP} up to a few tens of \unit{}{\micro\volt}, leading to Seebeck coefficients between \unit{200}{\micro\volt\per\kelvin} and \unit{1}{\milli\volt\per\kelvin}. These results make MTJs promising candidates for magnetic field controlled thermoelectric applications. In Al$_2$O$_3$ based MTJs the ratios of TMTP and of the tunneling magneto resistance (TMR) have been similar~\cite{Lin2012}. As a consequence, no spin dependent tunneling magneto thermocurrent (TMTC) has been observed \cite{Lin2012}. In MgO based MTJs, on the other hand, a significantly different amplitude of TMTP and TMR ratio has been measured \cite{Liebing2011,Walter2011b}. This should result in a spin dependence of the thermo electric current and hence in a TMTC. However, the TMTC of this important spintronic material has not been investigated experimentally so far.

Here, we experimentally study TMTP and TMTC in CoFeB/MgO/CoFeB MTJ nanopillars. The MTJs show a large TMTP ratio of up to \unit{55}{\%} and a TMR ratio up to  \unit{140}{\%}. Also the TMTC shows a clear spin dependence with an opposite sign and values down to \unit{-45}{\%}. These results can be well explained by the Onsager relation connecting thermal and electronic transport and indicate different tunneling thermal transport phenomena in MgO and Al$_2$O$_3$ based MTJs.

The experiments were carried out on nominally identical elliptical MTJ nanopillars with lateral dimensions of \unit{160}{\nano\metre}$\times$\unit{320}{\nano\metre} consisting of a synthetically pinned layer with a \unit{2}{\nano\metre} CoFeB top electrode, a \unit{1.5}{\nano\metre} MgO tunnel barrier and a \unit{3}{\nano\metre} CoFeB free layer. Details on the complete stack and on the fabrication process can be found elsewhere \cite{Liebing2011,Serrano-Guisan2008,SingulusCT}. All samples show TMR ratios between \unit{70}{\%} and \unit{140}{\%} with uni\-axial an\-iso\-tropy and single domain-like magnetization reversal. The coercive fields $\mu_0 H_\mathrm{K}$ range from \unit{8}{\milli\tesla} up to \unit{15}{\milli\tesla}.
\begin{figure}
  \centering
  \includegraphics[width=\textwidth]{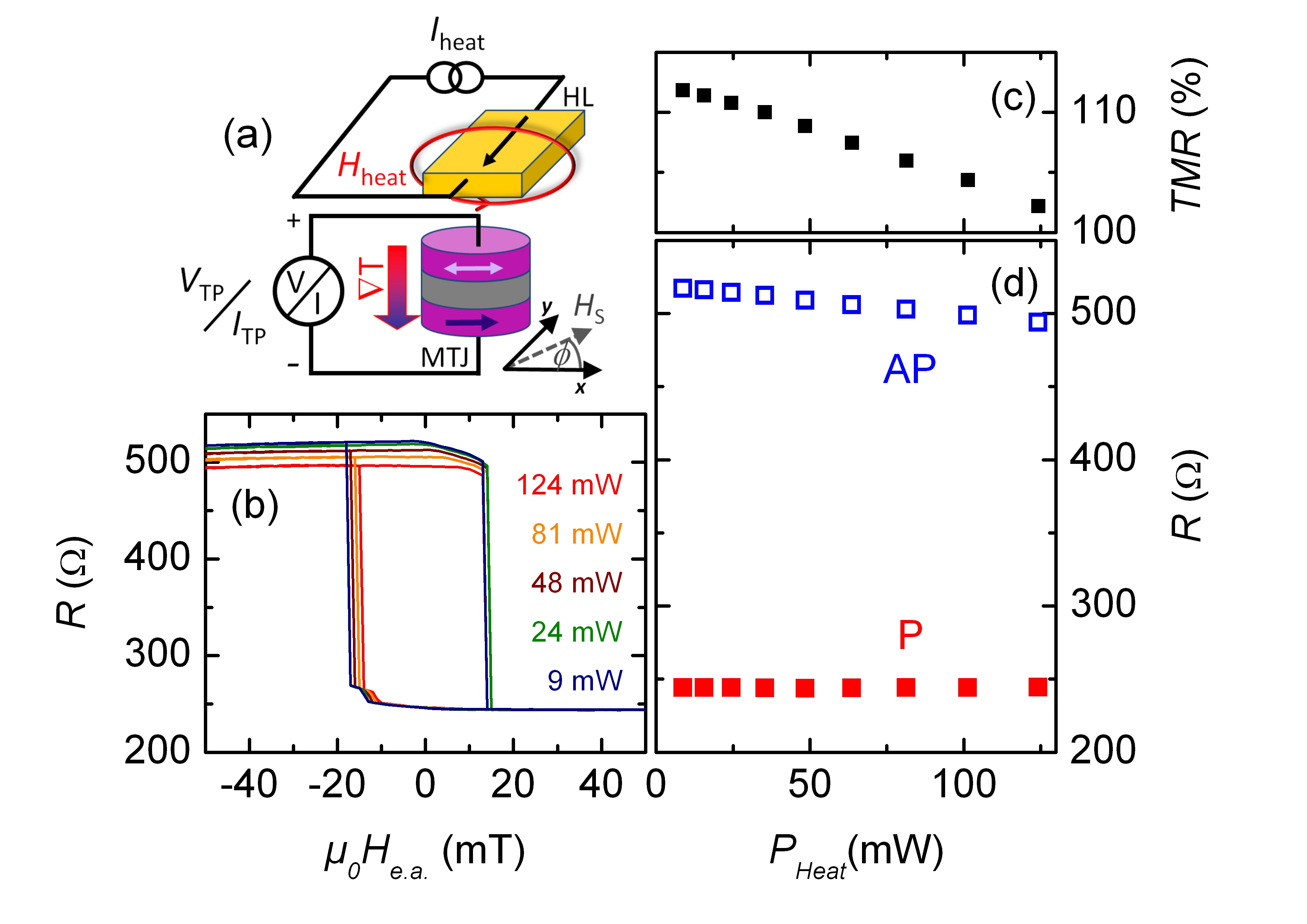}
  \caption{Power heating (\textit{P}\textsubscript{heat}) dependence of magneto electric transport properties of CoFeB/MgO based MTJ devices. (a) A schematic sketch of the experimental setup. (b) Free layer easy axis TMR hysteresis loops for different dc heating powers \textit{P}\textsubscript{heat}. (Note that \textit{H}\textsubscript{heat} is compensated.) (c) TMR as a function of \textit{P}\textsubscript{heat} derived from (b). (d) MTJ resistance $R_\mathrm{MTJ}$ in parallel (P, filled red squares) and antiparallel (AP, open blue squares) configuration as a function of \textit{P}\textsubscript{heat}.}
  \label{fig:Figure1}
\end{figure}

Thermal gradients are produced by ohmic heating (fig.~\ref{fig:Figure1}\,(a)) with a \unit{5}{\micro\metre} wide and \unit{70}{\nano\metre} thick gold heater line (HL) patterned on top of the MTJ and separated from the top contact of the nanopillar by a \unit{85}{\nano\metre} Ta$_2$O$_5$ dielectric.

DC heat currents up to \textit{I}\textsubscript{heat}=\unit{110}{\milli\ampere} were applied to the HL corresponding to a maximum power dissipation of \textit{P}\textsubscript{heat}=\unit{124}{\milli\watt}. In the presented measurements the effect of the Oerstedt field \textit{H}\textsubscript{heat} of \unit{0.1}{\milli\tesla} per \unit{}{\milli\ampere} applied heating current \textit{I}\textsubscript{heat} is always compensated by an appropriate external field.

\textit{V}\textsubscript{TP} is measured between the top contact (TC) and the bottom contact (BC) of the MTJ in an open electrical circuit using a HP\,334420A nanovoltmeter. Current measurements between TC and BC are carried out in a closed circuit using a Keithley\,6514 electrometer. To derive the pure thermally induced current (\textit{I}\textsubscript{TP}) in the presence of a temperature gradient first the spurious background current contribution across the MTJ (e.~g.~from rectification of unshielded ac signals) is measured for \textit{P}\textsubscript{heat}$=0$. Then the current for \textit{P}\textsubscript{heat}$\neq0$ is recorded and \textit{I}\textsubscript{TP} is derived by subtraction of the field loops for \textit{P}\textsubscript{heat}$\neq0$ and \textit{P}\textsubscript{heat}$=0$. Here a positive \textit{V}\textsubscript{TP} (\textit{I}\textsubscript{TP}) corresponds to a positive bias voltage (charge flow) between the TC and the BC. Both thermoelectric quantities are recorded for in-plane easy axis field sweeps between $\mu_0 H_\mathrm{e.a.}=\pm$\unit{50}{\milli\tesla}.

A reduction of the TMR ratio (fig. \ref{fig:Figure1}\,(c)) is observed when measuring the TMR for different heating powers (fig. \ref{fig:Figure1}\,(b)). This reduction is mainly caused by a decrease of the resistance in the antiparallel state ($R_\mathrm{AP}$), while the resistance in the parallel configuration ($R_\mathrm{P}$) remains stable (fig. \ref{fig:Figure1}\,(d)). This behavior is consistent with the decrease of the TMR ratio in MgO based MTJ structures with increasing temperature~\cite{Drewello2008,Hayakawa2006,Parkin2004,Yuasa2006}. To confirm this for our samples, the TMR was measured on a variable temperature probe station. A reduction of the TMR ratio by \unit{10}{\%} for an increase of the sample temperature from \unit{300}{\kelvin} to \unit{350}{\kelvin} was found in agreement with previous studies. We thus conclude that in our MTJs an applied \textit{P}\textsubscript{heat}$=\unit{125}{\milli\watt}$ leads to an increase of the average junction temperature of about \unit{50}{\kelvin}. Note that in our previous experiments samples with an approximately twice as thick dielectric layer between HL and TC were used~\cite{Liebing2011,Liebing2012}. There, no significant drop of the TMR was found for \textit{P}\textsubscript{heat} up to $P=\unit{60}{\milli\watt}$ speaking for a much weaker increase of the junction temperature.

\begin{figure}
  \centering
  \includegraphics[width=\textwidth]{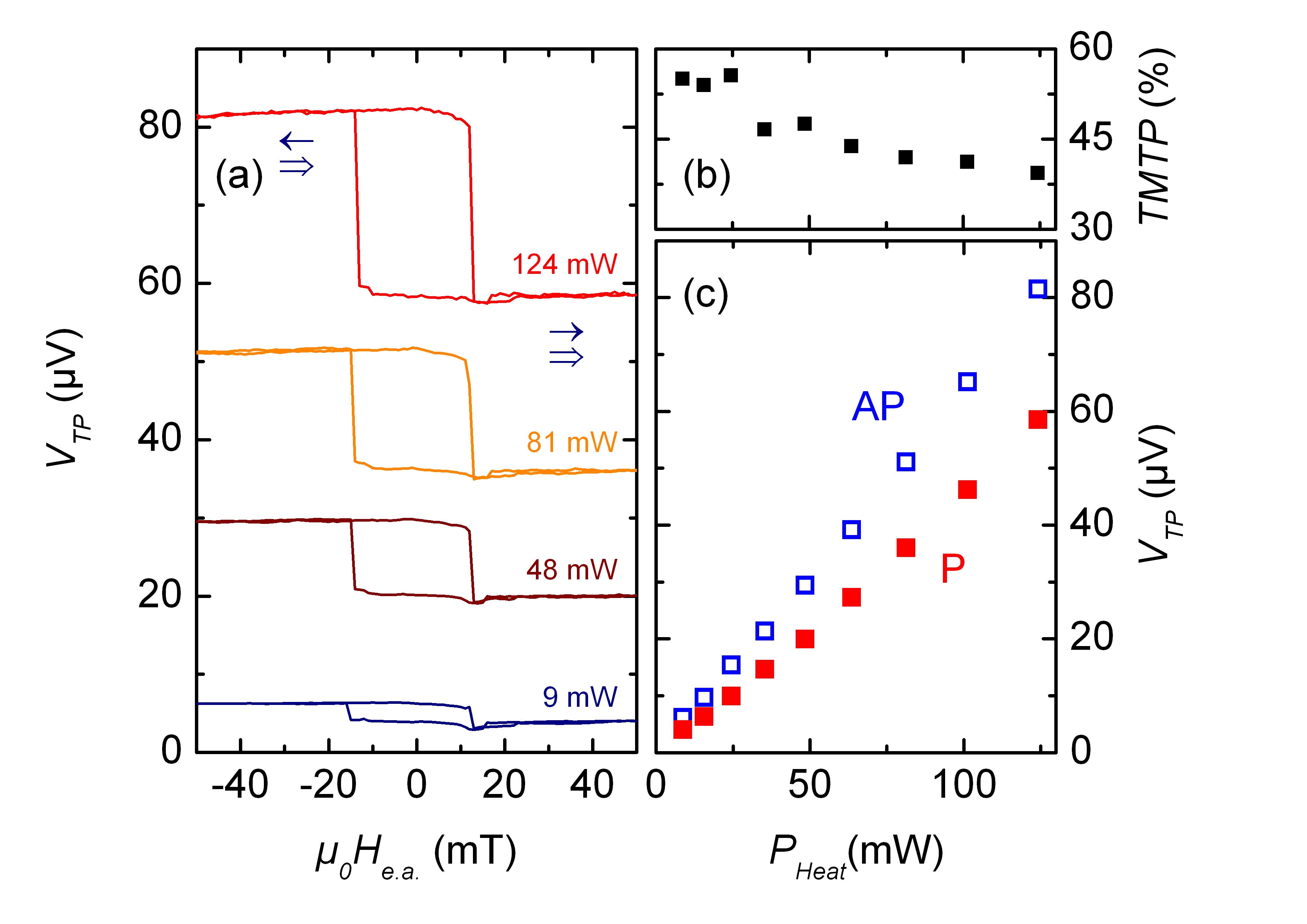}
  \caption{Thermal voltage \textit{V}\textsubscript{TP} measurements. (a) Easy axis TMTP loops under different applied dc heating powers \textit{P}\textsubscript{heat}. (b) TMTP ratios vs.~\textit{P}\textsubscript{heat} (c) \textit{V}\textsubscript{TP} as a function of \textit{P}\textsubscript{heat} for parallel (P, filled red squares) and antiparallel (AP, open blue squares) configuration of the MTJ.}
  \label{fig:Figure2}
\end{figure}

Figure \ref{fig:Figure2}\,(a) displays typical results for the thermopower \textit{V}\textsubscript{TP} measured under different dc heating powers (\textit{P}\textsubscript{heat}$=9,\,48\textrm{ and }\unit{124}{\milli\watt}$) as a function of the in-plane easy axis field $\mu_0 H_\mathrm{e.a.}$.
\textit{V}\textsubscript{TP} basically shows the same field dependence as the TMR with a hysteretic change of \textit{V}\textsubscript{TP} upon free layer magnetization reversal. In our experiment \textit{V}\textsubscript{TP} in the antiparallel state \textit{V}\textsubscript{TP}(AP) is always higher than \textit{V}\textsubscript{TP} in the parallel state \textit{V}\textsubscript{TP}(P). As displayed in figure \ref{fig:Figure2}\,(c) \textit{V}\textsubscript{TP}(AP) and \textit{V}\textsubscript{TP}(P) both scale almost linearly with the applied heat power up to the maximum applied \textit{P}\textsubscript{heat}$=\unit{124}{\milli\watt}$. Defining the TMTP ratio TMTP$=$(\textit{V}\textsubscript{TP}(AP)$-$\textit{V}\textsubscript{TP}(P))$/$\textit{V}\textsubscript{TP}(P) similar to the TMR we observe that the TMTP decreases from \unit{55}{\%} down to \unit{39}{\%} when increasing \textit{P}\textsubscript{heat} (fig. \ref{fig:Figure2}\,(b)). A temperature dependence of the TMTP ratio has also been predicted by ab initio calculations of the thermo electric power originating from the energy-dependent transmission probability trough the MgO barrier~\cite{Czerner2011}.

Based on a finite element modelling of the heat flow in our MTJs we can estimate the temperature drop across the MgO barrier of $\nabla T=\unit{45}{\milli\kelvin}$ at \textit{P}\textsubscript{heat}$=\unit{60}{\milli\watt}$\,~\cite{Liebing2011,Liebing2012}. Thus we can estimate the spin-dependent Seebeck coefficients for both parallel $S^\mathrm{P}_\mathrm{MTJ}=V_\mathrm{TP}(P)/\Delta T_\mathrm{MTJ}$ and antiparallel $S^\mathrm{AP}_\mathrm{MTJ}=V_\mathrm{TP}(AP)/\Delta T_\mathrm{MTJ}$ states \cite{Liebing2012}. Using the values of $S^\mathrm{P}_\mathrm{MTJ}$ and $S^\mathrm{AP}_\mathrm{MTJ}$ the value of $\Delta S_\mathrm{MTJ}\equiv S^\mathrm{AP}_\mathrm{MTJ}-S^\mathrm{P}_\mathrm{MTJ}=\unit{(240\pm110)}{\micro\volt\per\kelvin}$ is in good agreement with theoretically predicted spin-dependent Seebeck coefficients of $\Delta S_\mathrm{MTJ}=\unit{150}{\micro\volt\per\kelvin}$ for a comparable material system~\cite{Czerner2011}.
Note that the error of $\Delta S_\mathrm{MTJ}$ reflects the statistical uncertainty from averaging over the different samples and the uncertainty in the determination of the temperature gradient with numerically modelling~\cite{Liebing2012}.

\begin{figure}
  \centering
  \includegraphics[width=\textwidth]{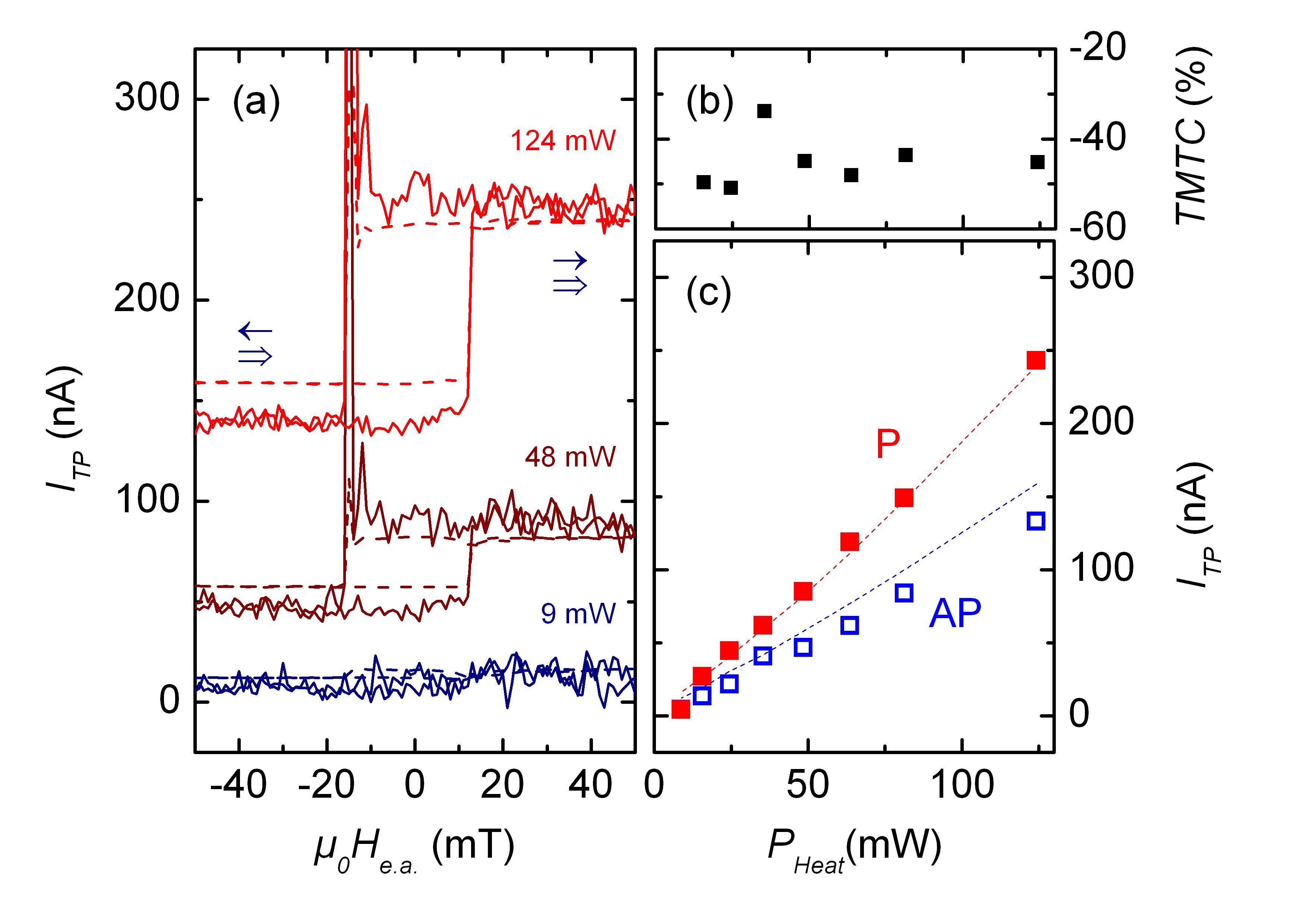}
  \caption{Thermal current \textit{I}\textsubscript{TP} measurements. (a) Easy axis TMTC loops under different applied dc heating powers \textit{P}\textsubscript{heat} (The solid lines correspond to the experimental data and the dashed lines to the Onsager model). (b) TMTC ratios vs.~\textit{P}\textsubscript{heat} (c) \textit{I}\textsubscript{TP} as a function of \textit{P}\textsubscript{heat} for parallel (P, filled red squares) and antiparallel (AP, open blue squares) configuration of the MTJ. The red dashed line and the blue dashed line correspond to calculated values using the Onsager relationship for the parallel and antiparallel configuration, respectively.}
  \label{fig:Figure3}
\end{figure}

Figure \ref{fig:Figure3}\,(a) shows the tunneling magneto thermocurrent \textit{I}\textsubscript{TP} measured in a closed circuit configuration as a function of the easy axis field for three different heating powers \textit{P}\textsubscript{heat}$=9,\,48,\unit{124}{\milli\watt}$ (solid lines in figure \ref{fig:Figure3}\,(a)). \textit{I}\textsubscript{TP} reveals similar hysteretic switching as TMR and \textit{V}\textsubscript{TP} but with opposite sign: \textit{I}\textsubscript{TP} is larger in the parallel than in the antiparallel state and for maximum \textit{P}\textsubscript{heat} \textit{I}\textsubscript{TP} nearly doubles upon magnetization reversal.
Note that especially in the P$\rightarrow$AP transition \textit{I}\textsubscript{TP} shows a significant overshoot. This overshoot is an artifact from the background subtration used to derive \textit{I}\textsubscript{TP} as described above. As shown in figure \ref{fig:Figure1}\,(b) a shift of the coercive field related to the P$\rightarrow$AP transition is observed when increasing \textit{P}\textsubscript{heat}. As \textit{I}\textsubscript{TP} is derived by subtration of two subsequently measured easy axis loops \textit{I(H)} for \textit{P}\textsubscript{heat}$\neq0$ and \textit{P}\textsubscript{heat}$=0$ the loop shift leads to incorrect values at the transition and hence to the observed overshoot in \textit{I}\textsubscript{TP}.

From the field sweep measurements in figure \ref{fig:Figure3}\,(a) we can derive the dependence of \textit{I}\textsubscript{TP}(P,AP) on \textit{P}\textsubscript{heat} presented in figure \ref{fig:Figure3}\,(c). \textit{I}\textsubscript{TP}(P) and \textit{I}\textsubscript{TP}(AP) again scale nearly linearly with \textit{P}\textsubscript{heat} and hence with the temperature gradient $\nabla$\textit{T}\textsubscript{MTJ} across the MTJ. In contrast to the behavior of \textit{V}\textsubscript{TP}, \textit{I}\textsubscript{TP}(AP) is smaller than \textit{I}\textsubscript{TP}(P). The TMTC ratio TMTC$=$(\textit{I}\textsubscript{TP}(AP)$-$\textit{I}\textsubscript{TP}(P))$/$\textit{I}\textsubscript{TP}(P) plotted in figure~\ref{fig:Figure3}\,(b) is negative at around \unit{-45}{\%} with no significant dependence on \textit{P}\textsubscript{heat}.

The experimental data can be well described by nonequilibrium thermodynamics \cite{Johnson1987,Johnson2010}.
According to the Onsager transport equations, the total electric current \textit{I} in the presence of the total voltage \textit{V} and a temperature gradient $\nabla$\textit{T}\textsubscript{MTJ} across the MTJ is given by
\begin{equation}
I=-\sigma V-\sigma S \nabla T_\mathrm{MTJ},
\label{equ:Onsager_relation}
\end{equation}
where $\sigma$ is the electrical conductance of the MTJ and \textit{S} is the Seebeck coefficient. Measurements of the junction resistance $R_\mathrm{MTJ}$ as well as \textit{V}\textsubscript{TP} and \textit{I}\textsubscript{TP} in an open circuit ($I=0$) and closed circuit ($V=0$) configuration, respectively, allow to determine $\sigma=(R_\mathrm{MTJ})^{-1}$ and $V_\mathrm{TP}=S\cdot\nabla T_\mathrm{MTJ}$. Using these values the predicted magnetic thermocurrent \textit{I}\textsubscript{TP} in closed circuit configuration can be computed as
\begin{equation}
I_\mathrm{TP}=-\frac{1}{R_\mathrm{MTJ}}V_\mathrm{TP}.
\end{equation}
The computed \textit{I}\textsubscript{TP} is plotted in figure~\ref{fig:Figure3}\,(a) and figure~\ref{fig:Figure3}\,(c) as   dashed lines. This simple model can well reproduces the main features of the experimental data. Note, however, that \textit{I}\textsubscript{TP} in the AP state is slightly lower than the predicted value and that for the P configuration a better agreement is found. As well as in the measurements an overshoot in the computed \textit{I}\textsubscript{TP} is found. This is consistent with the above explanation, as, again, the model data is derived from two measured easy axis loops at \textit{P}\textsubscript{heat}$=0$ ($R_\mathrm{MTJ}$) and \textit{P}\textsubscript{heat}$\neq0$ (\textit{V}\textsubscript{TP}).


In summary, we have presented a complete thermo\-electric characterisation of CoFeB/MgO/CoFeB based MTJ nanopillars in the presence of a thermal gradient. The magneto thermo\-power \textit{V}\textsubscript{TP} and the magneto thermo\-current \textit{I}\textsubscript{TP} both scale linearly with the applied thermal gradient and can be well described by the Onsager relations. \textit{V}\textsubscript{TP} and \textit{I}\textsubscript{TP} both show a hysteretic field dependence upon mag\-neti\-zation reversal but with opposite sign.

\textit{Note:} Recently, similar results in MgO based tunnel junctions based on optical heating have been observed \cite{Boehnke2013}.

We acknowledge funding by the DFG Priority Program SpinCaT, by EMRP JRP IND 08 MetMags and by EMRP JRP EXL-04 SpinCaT. The EMRP is jointly funded by the EMRP participating countries within EURAMET and the EU.

\end{document}